\documentstyle[12pt]{article}
\def\hybrid{\topmargin -20pt    \oddsidemargin 0pt
        \headheight 0pt \headsep 0pt
        \textwidth 6.25in       % A4 paper
        \textheight 9.5in       % A4 paper
        \marginparwidth .875in
        \parskip 5pt plus 1pt   \jot = 1.5ex}

\hybrid

\def\baselinestretch{1.2}

\catcode`\@=11

\def\marginnote#1{}
\newcount\hour
\newcount\minute
\newtoks\amorpm
\hour=\time\divide\hour by60
\minute=\time{\multiply\hour by60 \global\advance\minute by-\hour}
\edef\standardtime{{\ifnum\hour<12 \global\amorpm={am}%
        \else\global\amorpm={pm}\advance\hour by-12 \fi
        \ifnum\hour=0 \hour=12 \fi
        \number\hour:\ifnum\minute<10 0\fi\number\minute\the\amorpm}}
\edef\militarytime{\number\hour:\ifnum\minute<10 0\fi\number\minute}
\def\draftlabel#1{{\@bsphack\if@filesw {\let\thepage\relax
   \xdef\@gtempa{\write\@auxout{\string
      \newlabel{#1}{{\@currentlabel}{\thepage}}}}}\@gtempa
   \if@nobreak \ifvmode\nobreak\fi\fi\fi\@esphack}
        \gdef\@eqnlabel{#1}}
\def\@eqnlabel{}
\def\@vacuum{}
\def\draftmarginnote#1{\marginpar{\raggedright\scriptsize\tt#1}}

\def\draft{\oddsidemargin -.5truein
        \def\@oddfoot{\sl preliminary draft \hfil
        \rm\thepage\hfil\sl\today\quad\militarytime}
        \let\@evenfoot\@oddfoot \overfullrule 3pt
        \let\label=\draftlabel
        \let\marginnote=\draftmarginnote
   \def\@eqnnum{(\theequation)\rlap{\kern\marginparsep\tt\@eqnlabel}%
\global\let\@eqnlabel\@vacuum}  }

\def\preprint{\twocolumn\sloppy\flushbottom\parindent 2em
        \leftmargini 2em\leftmarginv .5em\leftmarginvi .5em
        \oddsidemargin -.5in    \evensidemargin -.5in
        \columnsep .4in \footheight 0pt
        \textwidth 10.in        \topmargin  -.4in
        \headheight 12pt \topskip .4in
        \textheight 6.9in \footskip 0pt
        \def\@oddhead{\thepage\hfil\addtocounter{page}{1}\thepage}
        \let\@evenhead\@oddhead \def\@oddfoot{} \def\@evenfoot{} }

\def\numberbysection{\@addtoreset{equation}{section}
        \def\theequation{\thesection.\arabic{equation}}}

\def\underline#1{\relax\ifmmode\@@underline#1\else
        $\@@underline{\hbox{#1}}$\relax\fi}

\def\titlepage{\@restonecolfalse\if@twocolumn\@restonecoltrue\onecolumn
     \else \newpage \fi \thispagestyle{empty}\c@page\z@
        \def\thefootnote{\fnsymbol{footnote}} }

\def\endtitlepage{\if@restonecol\twocolumn \else \newpage \fi
        \def\thefootnote{\arabic{footnote}}
        \setcounter{footnote}{0}}  %\c@footnote\z@ }

\catcode`@=12
\relax

\def\figcap{\section*{Figure Captions\markboth
        {FIGURECAPTIONS}{FIGURECAPTIONS}}\list
        {Figure \arabic{enumi}:\hfill}{\settowidth\labelwidth{Figure
999:}
        \leftmargin\labelwidth
        \advance\leftmargin\labelsep\usecounter{enumi}}}
 \relax
\def\tablecap{\section*{Table Captions\markboth
        {TABLECAPTIONS}{TABLECAPTIONS}}\list
        {Table \arabic{enumi}:\hfill}{\settowidth\labelwidth{Table
999:}
        \leftmargin\labelwidth
        \advance\leftmargin\labelsep\usecounter{enumi}}}
 \relax
\def\reflist{\section*{References\markboth
        {REFLIST}{REFLIST}}\list
        {[\arabic{enumi}]\hfill}{\settowidth\labelwidth{[999]}
        \leftmargin\labelwidth
        \advance\leftmargin\labelsep\usecounter{enumi}}}
 \relax
\makeatletter
\newcounter{pubctr}
\def\publist{\@ifnextchar[{\@publist}{\@@publist}}
\def\@publist[#1]{\list
        {[\arabic{pubctr}]\hfill}{\settowidth\labelwidth{[999]}
        \leftmargin\labelwidth
        \advance\leftmargin\labelsep
        \@nmbrlisttrue\def\@listctr{pubctr}
        \setcounter{pubctr}{#1}\addtocounter{pubctr}{-1}}}
\def\@@publist{\list
        {[\arabic{pubctr}]\hfill}{\settowidth\labelwidth{[999]}
        \leftmargin\labelwidth
        \advance\leftmargin\labelsep
        \@nmbrlisttrue\def\@listctr{pubctr}}}
 \relax
\makeatother
\newskip\humongous \humongous=0pt plus 1000pt minus 1000pt

\newif\ifdtup

\relax

\def\be{\begin{equation}}
\def\ee{\end{equation}}
\def\ba{\begin{eqnarray}}
\def\ea{\end{eqnarray}}

\def\no{\noindent}

\def\IR{\relax{\rm I\kern-.18em R}}

\def\IR{\relax{\rm I\kern-.18em R}}
\def\inv{^{\raise.15ex\hbox{${\scriptscriptstyle -}$}\kern-.05em 1}}

\begin{document}
%\draft

%\renewcommand{\theequation}{\arabic{equation}}
\renewcommand{\theequation}{\thesection.\arabic{equation}}

\newcommand{\beq}{\begin{equation}}
\newcommand{\eeq}[1]{\label{#1}\end{equation}}
\newcommand{\ber}{\begin{eqnarray}}
\newcommand{\eer}[1]{\label{#1}\end{eqnarray}}
\newcommand{\eqn}[1]{(\ref{#1})}
\begin{titlepage}
\begin{center}

\hfill hep--th/0205007\\
\hfill April 2002\\

\vskip .6in

{\large \bf Notes on Periodic Solitons}\footnote{Contribution to the 
proceedings of the RTN European network conference ``Quantum Structure of 
Space-time and the Geometric Nature of Fundamental Interactions" held in 
Corfu in September 2001; to be published in a special volume of 
{\em Fortschritte der Physik} edited by C. Kounnas, D. L\"ust and 
S. Theisen.}

\vskip 0.6in

{\bf Ioannis Bakas}\phantom{x}and\phantom{x} 
{\bf Christos Sourdis}
\vskip 0.1in
{\em Department of Physics, University of Patras, \\
GR-26500 Patras, Greece\\
{\tt bakas@ajax.physics.upatras.gr,\\
 sourdis@pythagoras.physics.upatras.gr}}\\
\vskip .2in

\end{center}

\vskip .8in

\centerline{\bf Abstract}

\no
We consider static solutions of the sine-Gordon theory defined on a cylinder, which 
can be either periodic or quasi-periodic in space. They are described by the different 
modes of a simple pendulum moving in an inverted effective potential and   
correspond to its libration or rotation. We review the decomposition 
of the solutions into an oscillatory sum of alternating kinks and anti-kinks or into
a monotonic train of kinks, respectively, using properties of elliptic functions. The 
two sectors are naturally related to each other by a modular transformation, whereas 
the underlying spectral curve of the model can be used to express the energy of the 
static configurations in terms of contour integrals \`a la Seiberg-Witten 
in either case. The stability properties are also examined by means of supersymmetric 
quantum mechanics, where we find that the unstable configurations are associated 
to singular superpotentials, thus allowing for negative modes in the spectrum of  
small fluctuations.

\end{titlepage}
\vfill
\eject

\def\baselinestretch{1.2}
\baselineskip 16 pt
\noindent
\section{Introduction}
\setcounter{equation}{0}
Solitons arise as static configurations of two-dimensional 
scalar field theories with potentials having more than one degenerate minima.   
In a typical situation, where the potential $U(\phi)$ is derived from a 
superpotential $W(\phi)$ that is conveniently normalized as 
$U(\phi) = 1/2 ({\partial}_{\phi} W(\phi))^2$, the energy functional of static
configurations is bounded from below \`a la Bogomol'nyi  
\be
E_{\rm stat} = \int_{-\infty}^{+\infty}  dx \left( {1 \over 2} 
({\phi}^{\prime} (x))^2 + U(\phi) 
\right) \geq \pm \left(W (+\infty) - W (-\infty) \right)   
\ee
as the space extends over $R$. 
The lower bound provides the energy of static configurations that satisfy 
the first order equations 
\be
{\phi}^{\prime} (x) = \pm {\partial W \over \partial \phi} ~, 
\ee
which also solve the second order equations of motion by construction. 
These are the defining  
equations for the solitons of the model provided that they extended 
between different degenerate vacua as $x$ ranges over space from $-\infty$ 
to $+\infty$, so that $W(+\infty) \neq W(-\infty)$. Kinks correspond to
solutions of the Bogomol'nyi equation with plus sign, whereas anti-kinks 
to minus sign. The notion also extends easily
to multi-component scalar field models in two space-time dimensions with
degenerate vacua. 

The construction of periodic solitons poses an interesting problem when the 
theory is defined on a two-dimensional cylinder $R \times S^1$, as  
the spatial dimension is compact and the Bogomol'nyi bound is identically zero.     
Therefore, if one is interested in finding periodic solutions of the 
first order equation ${\phi}^{\prime} (x) = \pm {\partial}_{\phi} W (\phi)$ 
(or its generalizations for multi-component models), there will be no 
choice of the integration constants that amounts to non-trivial solutions 
of the same equations having a periodic structure, $\phi(x + L) = \phi(x)$, 
other than the constant 
configurations sitting at the bottom of the potential 
with zero energy. Recall also that the 
physical meaning of the integration constants of Bogomol'nyi equations is 
provided by the location of the  
solitons in $x$, as it is done in non-compact space 
once and for all, and leaves no room for the size of the space itself,  
i.e., its period L, to fit into the available moduli. Following this simple 
observation, it turns out that the notion of periodic solitons is ill-defined 
for the class of two-dimensional scalar models under consideration, as no 
such solutions seem to be possible. We also note for  
completeness that in the presence of discrete symmetries in the model, for
instance $Z_2$-invariance under $\phi \rightarrow -\phi$, one might also 
consider configurations with twisted boundary conditions over $S^1$, 
$\phi(x + L) = - \phi(x)$. Furthermore, for models with non-trivial topology 
in field space, it is also possible to have configurations with non-trivial winding 
number over $S^1$, $\phi(x + L) = \phi(x) + 2\pi$ and so on. It is straightforward to 
see that the no-go theorem stated above also extends to such topological sectors 
and hence the existence of periodic solutions of Bogomol'nyi equations remains 
impossible in its generality.  

There are two possibilities one might subsequently entertain in order to make  
the notion of ``periodic solitons" more precise and meaningful. The first one is 
to shift the vacuum energy density of the theory by an appropriately chosen 
constant so that the modified potential $\tilde{U}(\phi)$ can be derived 
from a new superpotential $\tilde{W}(\phi)$, which is not single-valued     
but its derivatives are single-valued functions. This results into a modified
form of Bogomol'nyi equations, namely 
${\phi}^{\prime} (x) = \pm {\partial}_{\phi} \tilde{W}(\phi)$, which are  
still first order and can support non-trivial periodic solutions that also satisfy 
the same second order classical equations of motion of the 
theory by construction. According to
this possibility, the completion of the perfect square term in the energy functional 
of static configurations is done differently for theories defined over compact 
space in order for the boundary term to compute their energy correctly. Note, however, 
there is no guarantee that the periodic solitons defined in this fashion are
always stable under small fluctuations, as for the case of ordinary solitons on $R$.  
The second possibility is to rely on the integrability properties of many 
two-dimensional field theory models that admit solitons on the real line and device 
a suitable superposition principle for constructing exact solutions 
by summing up trains of ordinary kinks or anti-kinks whose centers  
are equally spaced on the real line using a uniform length scale $L$. Despite 
the non-linearity of the models, such superposition principle is always possible
in integrable systems and leads to periodic solutions of the second order 
classical equations of motion that can also be named periodic solitons by their 
construction. 

Although these two possibilities appear to be different in nature, 
they turn out to be equivalent in several field theory models of current interest
like sine-Gordon,  $\lambda \phi^4$, etc that possess degenerate vacua, thus 
leading to an unambigious definition of the notion of periodic solitons in such
two-dimensional models. Here, we will 
only review some aspects of the construction and the basic properties of 
periodic solitons arising in the sine-Gordon model and leave further details
and generalizations to future publications \cite{ba}. We will also spell out some 
connections with the Seiberg-Witten theory of BPS states in four-dimensional 
supersymmetric gauge theories as application of our results.  
  
\section{Sine-Gordon model on $R \times S^1$}
\setcounter{equation}{0}
Consider the sine-Gordon theory with Lagrangian density 
\be
{\cal L} = {1 \over 2} (\partial_\mu \phi)(\partial^\mu \phi) - U(\phi) ~; 
~~~~ U(\phi) = 1 + {\rm cos} \phi  
\ee
defined on a two-dimensional cylindrical space-time $R \times S^1$.  
We are interested in the construction of static configurations satisfying the
classical equations of motion 
\be
{\phi}^{\prime \prime} (x) = {\partial U \over \partial \phi} 
\ee
with strictly periodic boundary conditions on the spatial direction 
\be
\phi(x+L) = \phi (x) 
\ee
with period $L$; later we will also take advantage of the non-trivial topology 
in the field space of the model and consider solutions in the winding sector, 
i.e., quasi-periodic solutions with $\phi(x+L) = \phi(x) + 2\pi$. 
We proceed by multiplying both sides of the equation of motion with 
${\phi}^{\prime}(x)$, which upon integration yields
\be
{1 \over 2} \left({\phi}^{\prime} (x)\right)^2 - U(\phi) = -C ~, 
\ee
where $C$ denotes the corresponding integration constant. We can think of this
equation as describing a constant ``energy" condition for a point particle that
moves in an inverted potential $-U(\phi)$ with respect to an effective ``time"
$x$ \cite{ma}. 
This is an appropriate point of view for constraining the values of $C$
associated with strictly periodic solutions as they correspond to bounded 
effective motion, thus leading to $-2 \leq -C \leq 0$; later we will also let
$-C \geq 0$ to account for quasi-periodic solutions in the winding sector of 
the model that correspond to unbounded effective motion. In the language of a
classical pendulum, which is natural to consider for the sine-Gordon potential, 
these two different cases correspond to the modes of libration and rotation, 
respectively.

It is convenient to parametrize for the present purposes the parameter $C$ as
\be
C=2(1-k^2) ~; ~~~~ 0 \leq k \leq 1 ~,  
\ee
whereas $k^{\prime}$ will denote later the complementary modulus, 
$k^2 + {k^{\prime}}^2 = 1$. Then, one easily finds the static solution
\be
\phi(x) = \pm 2{\rm sin}^{-1} \left(k{\rm sn}(x + x_0; k) \right) 
\ee
with $L$ being identified with the real period of the Jacobi elliptic function
${\rm sn}(x+L; k) = {\rm sn}(x, k)$, i.e., 
\be
L = 4mK(k) ~; ~~~~ m ~~ {\rm positive} ~~ {\rm integer}, 
\ee
where $k$ serves as the modulus of the associated complete elliptic integral 
of the first kind, $K(k)$. The two possible signs arise from taking the
square root before integrating the resulting first order equation and they are
compatible with the discrete symmetry $\phi \rightarrow -\phi$ of the 
sine-Gordon model. Also, the integration constant $x_0$ determines the location 
of the configuration and can be set equal to zero without loss of generality.  

It is also interesting to consider two special 
limits that arise for $k=1$ (the point particle being at the verge of 
becoming unbounded) and $k=0$ (the point particle sitting at the 
bottom of the effective potential $-U(\phi)$):     

\noindent
\underline{$k=1$}: 
in this case we have ${\rm sn}(x; k=1) = {\rm tanh}x$ and therefore
the static solution (with plus sign) becomes 
\be
\phi(x) = 2 {\rm sin}^{-1}({\rm tanh}x) \equiv 4 {\rm tan}^{-1} e^x - \pi ~,
\ee
which is the usual kink solution centered at $x=0$ ($x_0$ more generally) 
that interpolates monotonically
between $\phi = -\pi$ for $x \rightarrow -\infty$ and $\phi = \pi$ for 
$x \rightarrow +\infty$ in the sine-Gordon potential; 
in this case the spatial dimension $S^1$ decompactifies 
to $R$ as $K(1)$, and hence the period,  becomes infinite. 
Note that the anti-kink solution will emerge in this limit, if we take
the opposite sign in the general periodic expression.   

\noindent
\underline{$k=0$}: 
in this case we have ${\rm sn}(x; k=0) = {\rm sin}x$ and therefore
we simply obtain the sphaleron configuration
\be
\phi(x) = 0 ~,
\ee
which represents a constant configuaration at the top of the sine-Gordon potential 
$U(\phi)$. 

Thus, for generic values of the modulus $0 \leq k \leq 1$, we have a periodic  
solution whose particular properties are worth studying, as they extrapolate from 
the stable structure of a monotonic kink to the highly unstable structure 
of a sphaleron. We will also learn some lessons from this study later,  
by using two different methods based on the direct 
computation of the fluctuation spectrum as well as on methods of 
supersymmetric quantum mechanics.  

Next, we illustrate a superposition principle underlying the periodic
solution of the non-linear 
sine-Gordon equation for generic values of $k$. Consider the
usual kink solutions of the model written in a convenient equivalent form 
\be
\phi_{\pm} (x) = 2i {\rm log} {1 \mp i e^x \over 1 \pm i e^x} - \pi ~, 
\ee
where $\pm$ correspond to a kink or an anti-kink configuration on $R$.  
Then, using standard identities from the theory of Jacobi elliptic functions, 
it is fairly straightforward to decompose the static periodic solution 
$\phi(x) = 2{\rm sin}^{-1} (k {\rm sn}(x; k))$ as follows:
\ba
\phi(x) & = & \sum_{n= -\infty}^{+\infty} \left( 2i {\rm log} 
{1 -i {\rm exp}\left({\pi \over 2K(k^{\prime})}  
\left(x-4nK(k)\right)\right)
\over 
1 +i {\rm exp}\left({\pi \over 2K(k^{\prime})}  
\left(x-4nK(k)\right)\right)} - \pi \right) + \nonumber\\
& + & \sum_{n= -\infty}^{+\infty} \left( 2i {\rm log} 
{1 +i {\rm exp}\left({\pi \over 2K(k^{\prime})}  
\left(\left(x+2K(k)\right) -4nK(k)\right)\right)
\over 
1 -i {\rm exp}\left({\pi \over 2K(k^{\prime})}  
\left(\left(x+2K(k)\right) -4nK(k)\right)\right)} - \pi \right) . 
\ea
Therefore, the periodic solution is composed of an alternating sum of kinks 
and anti-kinks separated from each other by $2K(k)$, which produce an 
oscillatory train solution with net period $4K(k)$ (see also \cite{fo}).  

\section{Spectrum of small fluctuations}
\setcounter{equation}{0}

The stability properties of a classical solution can be investigated in general  
by studying the spectrum of small fluctuations around it. It is well known
that the relevant Schr\"odinger problem for the spectrum is given by
\be
\left(-{d^2 \over dx^2} + V(x) \right) \psi_k (x) = {\omega_k}^2 
\psi_k (x) ~, 
\ee
where the Schr\"odinger potential is  
\be
V(x) = {\partial^2 U \over \partial \phi^2} \mid_{\phi_{\rm cl}(x)} 
\ee
for any given classical field configuration $\phi_{\rm cl}(x)$. 
Thus, stability is ensured by the absence of
tachyonic (negative) modes. 
The spectrum can be computed directly, when this is 
possible, but in many typical 
situations the Schr\"odinger potential assumes a form as in 
supersymmetric quantum mechanics, i.e., 
\be
V(x) = {\cal W}^2 (x) - {\cal W}^{\prime} (x) ~, 
\ee
where ${\cal W}(x)$ is the corresponding superpotential \cite{co}. In these cases
one concludes automatically that the spectrum is bounded from below by zero, 
thus establishing the desired stability properties without any further 
computation. There are some exceptional 
circumstances, however, where such conclusions should be drawn with care
in order to avoid wrong statements. 

The static periodic solution of the sine-Gordon equation is expected to be
unstable because of its oscillatory nature, unlike the stability of 
ordinary kinks which are monotonic functions. Direct computation of the
spectrum in this case amounts to solving the Schr\"odinger problem with 
potential
\be
V(x) = 2k^2 {\rm sn}^2 (x ; k) -1 ~,  
\ee
known as the Lam\'e potential. It can be easily verified that it supports
the following three square-integrable solutions:
\ba
\psi(x) & = & {\rm cn} (x ; k) ~, ~~~~ {\rm with} ~~ \omega^2 = 0 ~, 
\nonumber\\
\psi(x) & = & {\rm dn} (x ; k) ~, ~~~~ {\rm with} ~~ \omega^2 = k^2 - 1 ~,\\
\psi(x) & = & {\rm sn} (x ; k) ~, ~~~~ {\rm with} ~~ \omega^2 = k^2 . 
\nonumber 
\ea
The second solution is clearly tachyonic for $0 \leq k < 1$, which proves
the instability of the given solution in agreement with our intuition 
(see also \cite{ma}). 

On the other hand, the Lam\'e 
potential can be derived from a superpotential, as in supersymmetric 
quantum mechanics, with
\be
{\cal W}_k(x) = {{\rm dn}(x ; k) {\rm sn}(x ; k) \over {\rm cn}(x ; k)} ~.  
\ee
Therefore, one would expect to have no negative modes at all by relying 
on the general properties of supersymmetric quantum mechanics; this,   
in turn, poses an interesting puzzle by itself of having instability versus 
supersymmetry. The aparent confict is enhanced in the 
limit $k=0$, where it becomes obvious, as we have
\be
V_{k=0}(x) = -1 ~, ~~~~ {\cal W}_{k=0}(x) = {\rm tan}(x) ~.  
\ee
In the other limit, where $k=1$, we have
\be
V_{k=1}(x) = 1 -{2 \over {\rm cosh}^2 x} ~, ~~~~ {\cal W}_{k=1}(x) 
= {\rm tanh}(x) 
\ee
and the conflict goes away; note at this end that the supersymmetric partner 
of $V_{k=1}(x)$ is the constant potential $V(x) = 1$ with manifestly 
positive spectrum, as it is required for an ordinary kink on $R$.  
   
The resolution to this paradox is provided by the simple fact that the
superpotential ${\cal W}_k(x)$ is not regular everywhere in its domain
of definition, for all $0 \leq k < 1$. 
It is sufficient to note for the present problem that
the elliptic function ${\rm cn}(x ; k)$ develops zeros for all real  
$x = (2m +1)K(k)$ and hence the corresponding ${\cal W}_k(x)$ becomes
singular at two points, as $x$ ranges from 0 to $4K(k)$; for ${\cal W}_{k=0}(x)$,  
in particular, the singularity occurs at $x = \pi /2$ 
and $3\pi /2$, as $x$ 
ranges from 0 to $ 4K(0) = 2 \pi$. It has been known for some time that the
bear facts of supersymmetric quantum mechanics should be used with care under
such circumstances, for  
the case of singular superpotentials could obstruct their validity, like the
positivity of the spectrum (see \cite{co} and references therein). 
Finally, we note for completeness that the 
static solution in the twisted sector of the model, which has anti-periodic 
boundary conditions  
$\phi(x + 2K(k)) = - \phi(x)$, exhibits stability as the troublesome negative
mode $\psi(x) = {\rm dn} (x; k)$ has real period $2K(k)$ and hence is moded 
out by the twisting \cite{is}. 
Put it differently, in the language of supersymmetric 
quantum mechanics, the singularities of the superpotential occuring at 
$x = K(k)$ and $3K(k)$ are removed from the interval obtained by twisting, 
as they constitute the fixed points of the $Z_2$ action.  

\section{Spectral curve and its applications}
\setcounter{equation}{0}

Next, we provide a natural algebro-geometric description of the energy of periodic 
static configurations of the sine-Gordon equation, 
which is reminiscent of the Seiberg-Witten expression for the mass of the BPS  
states in $N=2$ supersymmetric Yang-Mills theories for the group $SU(2)$
\cite{sw}. Along these  
lines we will also examine the behaviour of static solutions under the action of modular
transformations of an underlying Riemann surface and show that the two different 
modes of evolution, namely the libration and rotation of a 
classical pendulum, are intimately connected
to each other by them. Actually, this is not a surprise, as the 
underlying integrable system of Seiberg-Witten theory is the sine-Gordon model 
\cite{mw}, whereas the supersymmetric BPS states of four-dimensional gauge theory are
monopoles and dyons.  

It is sufficient for this purpose to consider the zero curvature 
formulation of the static sine-Gordon equation in terms of the Nahm equations
\be
{dT_{i}(s) \over ds} = {1 \over 2} \sum_{j,k=1}^{3} 
\epsilon_{ijk} [T_{j}(s) , ~ T_{k}(s)] ~; 
~~~~ i=1, 2, 3 
\ee
for the special choice of matrices
\be
T_1 = -2i \left({\rm cos}{1 \over 2} \phi\right) \sigma_1 ~, ~~~ 
T_2 = 2 \left({\rm sin}{1 \over 2} \phi\right) \sigma_2 ~, ~~~ 
T_3 = {1 \over 2} \left({d \phi \over ds} \right) \sigma_3 ~, 
\ee
where $\sigma_i$ are the three Pauli matrices. The system of Nahm equations 
admits the zero curvature formulation
\be
[L(s) , ~ {d \over ds} + M(s)] = 0 ~,  
\ee
where
\ba
L & = & T_1 + i T_2 -2i \zeta T_3 + {\zeta}^2 (T_1 - i T_2) ~,\nonumber\\
M & = & -i T_3 + \zeta (T_1 - i T_2) 
\ea
using a spectral parameter $\zeta$. Note that the choice of Nahm matrices 
above yields the sine-Gordon equation for the variable $x = 2i s$, 
\be
{\phi}^{\prime \prime} (x) + {\rm sin} \phi(x) = 0 ~, 
\ee
as required.    

Let $\eta$ denote the eigenvalue of the Lax matrix $L(s)$, which is defined 
by the equation
\be
{\rm det} \left(L(s) - \eta 1 \right) = 0 
\ee
that yields the quartic algebraic equation in the complex variable $\zeta$,
\be
{\eta}^2 + 1 + 2 {\zeta}^2 \left( -{1 \over 2} 
\left({\phi}^{\prime}(x)\right)^2 + {\rm cos} \phi(x) \right) + 
{\zeta}^4 = 0 ~.  
\ee
Since the Lax pair formulation of the problem implies an isospectral flow for 
the matrix $L$, the coefficients of the resulting polynomial in $\zeta$ have
to be independent of $x$, thus providing the conservation laws of the system. 
In our case, using the notation adopted earlier, we have
\be
-{1 \over 2} \left({\phi}^{\prime}(x)\right)^2 + {\rm cos} \phi (x) = C-1 
\equiv 1-2k^2 ~. 
\ee
This defines the specral curve of the system, which is a Riemann surface of
genus 1 (see also \cite{st}). 

For bounded motion in the inverted sine-Gordon potential $-U(\phi)$, i.e., 
for $0 \leq k < 1$, the four roots of the quartic polynomial in $\zeta$ 
are all complex occuring in conjugate pairs 
\be
k+ ik^{\prime} , ~~~ k-ik^{\prime} , ~~~ -(k+ik^{\prime}) , ~~~ 
-(k-ik^{\prime}) ~. 
\ee
Then, transforming to the Weierstrass normal form of the spectral curve,  
\be
y^2 = 4(x-e_1)(x-e_2)(x-e_3) ~; ~~~~ \sum_{i=1}^3 e_i = 0 ~, 
\ee
we find the three roots $e_1 \geq e_2 \geq e_3$ to be given by
\be
e_1 = {1 \over 3}(2-k^2) , ~~~ e_2 = {1 \over 3} (2k^2 -1) , ~~~ 
e_3 = -{1 \over 3} (1+k^2) ~. 
\ee

The energy of the periodic static configuration can be computed directed in 
sine-Gordon theory. It is convenient to shift the vacuum 
energy density by the constant $C$, thus defining a new potential 
$\tilde{U}(\phi) = U(\phi) - C$, which does not alter the classical 
equations of motion.  
Then, it can be easily seen that the energy of our periodic static 
configuration can be cast into the form  
\be
E_{{\rm stat}}(k) = 8 \oint_a {dx \over y} 
(x-e_2) ~,  
\ee
using a contour integral 
over the $a$-cycle of the corresponding spectral curve, which encircles the 
branch cut between the roots $e_2$ and $e_3$. If we had taken the static 
solution in the twisted sector of the model with 
$\phi(x + 2 K(k)) = - \phi(x)$, the corresponding energy would have been 
half of it. Also note in this context 
that when the $a$-cycle shrinks to zero size (i.e., when $e_2 \rightarrow e_3$ 
or else $k \rightarrow 0$), one obtains the normalized energy of the sphaleron
configuration sitting at the top of the potential; likewise, when the $b$-cycle 
shrinks to zero size (i.e., when $e_1 \rightarrow e_2$ or else $k \rightarrow 1$),  
one obtains the energy of the usual kink and anti-kink configurations in 
non-compact space.  

At this point it is natural to consider the case of unbounded motion in more
detail, which defines solutions in the winding sector of the sine-Gordon 
model that correspond to the rotational mode of a classical pendulum, as 
opposed to the strictly periodic mode of libration. It is sufficient for
this purpose to consider values of the modulus $k$ bigger than 1, in which
case $-C \geq 0$. The extension to those values is naturally defined by 
the following transformation of the modulus $k$ and its complementary 
value $k^{\prime}$, 
\be
\tilde{k} = {1 \over k} ~, ~~~~ {\tilde{k}}^{\prime} = 
i {k^{\prime} \over k} ~,  
\ee
which results from the transformation law of the complex modulus $\tau$,  
\be
\tilde{\tau} = {\tau \over \tau + 1} ~; ~~~ {\rm where} ~~ 
\tau = i{K(k^{\prime}) 
\over K(k)} ~.  
\ee
This corresponds to the modular transformation $TST$, which is composed from
the elementary modular transformations $T: \tau \rightarrow \tau + 1$ and 
$S: \tau \rightarrow -1/\tau$. As such, it amounts to a transformation of
the complete elliptic integrals of the first kind, given by, 
\be
K(\tilde{k}) = k \left(K(k) + i K(k^{\prime}) \right) , ~~~~ 
K({\tilde{k}}^{\prime}) = k K(k^{\prime})  
\ee
and turns the four complex roots of the quartic polynomial of the spectral 
parameter $\zeta$ into four real roots, respectively,  
\be
\pm (\tilde{k} \pm i {\tilde{k}}^{\prime} ) = \pm \left({1 \over k} \mp 
{k^{\prime} \over k} \right) . 
\ee
It can be readily seen that under the modular transformation 
$\tau \rightarrow \tilde{\tau}$, we have
\be
{\rm exp} \left({\pi \over 2 K(k^{\prime})} (x - 2n K(k))\right) \rightarrow 
(-1)^n {\rm exp} \left({\pi \over 2 K(k^{\prime})} 
\left({x \over k} - 2n K(k)\right) \right) 
\ee
and so the oscillatory train of alternating kinks and anti-kinks, with
period $4K(k)$, turns into
a monotonic superposition of kinks (or anti-kinks) only.      

In effect, the Dehn twist prescribed on the torus of the corresponding
Jacobi variety, where $x$ takes real values along a closed cycle,  
maps strictly periodic solutions into solutions belonging in the winding 
sector of the sine-Gordon model with real quasi-period $2kK(k)$,  
\be
\phi(x + 2kK(k)) = \phi(x) + 2 \pi ~. 
\ee
In fact, using the identity 
$\tilde{k} {\rm sn}(x; \tilde{k}) = {\rm sn}(x/k ; k)$, we can easily  
cast the resulting monotonic solution of superimposed kinks 
in closed form and verify directly that 
under a $2kK(k)$-shift we obtain the desired quasi-periodic behaviour. 
Then, it turns out that the energy of this configuration can be 
written in the form
\be
\tilde{E}_{\rm stat} (k) = 4 \left( \oint_a 
{dx \over y} (x-e_2) - \oint_b {dx \over y} (x-e_2) \right) ,
\ee
using contour integration over the $a$- and $b$-cycles of the Riemann 
surface with complex modulus $\tau$ having $k>1$. Of course, as before, the
result is conveniently presented after shifting the vacuum energy density  
of the model by the constant $C$.
    
\section{Conclusions}
\setcounter{equation}{0}

In summary, we derived a classic and intuitive result effectively stating that 
the libration of a pendulum can be decomposed into an alternating train
of kinks and anti-kinks, whereas its rotational motion can be decomposed into
a train of kinks or anti-kinks only, depending on the  
two possible orientations of rotation.  
The monotonic nature of the solution in the latter case, also  
ensures the absence of any negative eigenvalues from the spectrum of
small fluctuations around it; clearly, the 
troublesome eigenvalue $k^2 -1$ of the 
corresponding Lam\'e potential ceases to be tachyonic for $k \geq 1$, thus
leading to stability. Put it differently, the superpotential ${\cal W}_k(x)$ of
the associated problem in supersymmetric quantum mechanics becomes 
non-singular in its domain of definition after a modular transformation 
and, hence, the eigenvalues turn out to be 
bounded from below by zero. Finally, we note that the expressions for the 
energy of static configurations in the twisted or in the winding sector of
the sine-Gordon theory coincide with the mass formula of certain BPS states 
in four-dimensional Seiberg-Witten theory; further details will be presented
elsewhere \cite{ba}.

\vskip.4in
\centerline{\bf Acknowledgements}
\noindent
This work was supported in part by the European research and training networks
``Superstring Theory" (HPRN-CT-2000-00122) and ``The Quantum Structure of 
Space-time" (HPRN-CT-2000-00131). The research of C.S. is also supported by
a graduate student fellowship ``C. Caratheodory" under the grant number 2453  
from the University of
Patras, which we also gratefully acknowledge.  
%%%%%%%%%%%%%%%%%%%%%%%%%%%%%%%%%%%%%%%%%%%%%%%%%%%%%%%%%%%%%%%%%%%%

\end{document}